\documentclass[12pt,epsf]{article}

\usepackage{graphicx}

\newcommand{\be}{\begin{equation}}

\newcommand{\ee}{\end{equation}}

\newcommand{\ba}{\begin{array}}

\newcommand{\ea}{\end{array}}

\begin{document}
\begin{titlepage}
\vspace{.5in}
\begin{flushright}
CQUeST-2007-0145
\end{flushright}
\vspace{0.5cm}

\begin{center}
{\Large\bf Dynamics of false vacuum bubbles with nonminimal coupling}\\
\vspace{.4in}

  {$\rm{Bum-Hoon \,\, Lee}^{\dag\P}$}\footnote{\it email:bhl@sogang.ac.kr}\,\,
  {$\rm{Chul \,\, H. \,\, Lee}^{\ddag}$}\footnote{\it email:chulhoon@hanyang.ac.kr}\,\,
  {$\rm{Wonwoo \,\, Lee}^{\S}$}\footnote{\it email:warrior@sogang.ac.kr}\,\,
  {$\rm{Siyoung \,\, Nam}^{\P}$}\footnote{\it email:stringphy@gmail.com}\,\,
  {$\rm{Chanyong \,\, Park}^{\P}$}\footnote{\it email:cyong21@sogang.ac.kr}\\

  {\small \dag \it Department of Physics, Sogang University, Seoul 121-742,
  Korea}\\
  {\small \S \it Research Institute for Basic Science, Sogang University, Seoul 121-742,
  Korea}\\
  {\small \P \it Center for Quantum Spacetime, Sogang University, Seoul 121-742,
  Korea}\\
  {\small \ddag \it  Department of Physics, and BK21 Division of Advanced Research and Education in Physics, Hanyang University, Seoul 133-791,
  Korea}\\

\vspace{.5in}
\end{center}
\begin{center}
{\large\bf Abstract}
\end{center}
\begin{center}
\begin{minipage}{4.75in}

{\small \,\,\,\, We study the dynamics of false vacuum bubbles. A
nonminimally coupled scalar field gives rise to the effect of
negative tension. The mass of a false vacuum bubble from outside
observer's point of view can be positive, zero, or negative. The
interior false vacuum has de Sitter geometry while the exterior
true vacuum background can have geometry depending on the vacuum
energy. We show that there exist expanding false vacuum bubbles
without the initial singularity in the past.}


\end{minipage}
\end{center}
\end{titlepage}

\newpage
\section{ Introduction \label{sec1}}

Can a false vacuum bubble expand within the true vacuum
background? Or, is an expanding false vacuum bubble always inside
the horizon of a black hole from outside observer's point of view?
If there was a dynamical spacetime foam structure in the very
early universe \cite{wheeler}, the detail structure and evolution
will depend on the cosmological constant. In the context of a
bubble nucleation and dynamics, this phenomenon may be described
as follows. A true vacuum bubble can always be nucleated somewhere
within the false vacuum background as well as a false vacuum
bubble nucleated within the true vacuum bubble background. Some
bubbles may expand while some bubbles collapse. Some of them may
be connected by wormholes. Then the whole spacetime may have the
complicated vacuum or spacetime structure due to the above
processes. Earlier works \cite{blau01, blau02, aguirre1,
freivogel1} show that the unbound solution representing expanding
false vacuum bubble does not exist because the solution for the
junction equation can not cover all ranges of $r$. To obtain an
expanding false vacuum bubble, the mass of the bubble should be
over some critical value. To the observer in the exterior
spacetime the expanding false vacuum bubble will be inside the
black hole horizon. Only to the observer inside bubble will it
appear to expand from a very small size to infinity. However,
these bubbles start from an initial singularity. Moreover, there
is a puzzle that the entropy of the expanding false vacuum region
is greater than the entropy of the black hole surrounding it
\cite{freivogel1, banks, entro01}. Are these descriptions always
true? With nonminimal coupling we will show that the unbound
solutions representing expanding false vacuum bubbles can exist.
On the other hand the idea of the string theory landscape has a
vast number of metastable vacua \cite{susskind}. One of the
intriguing features of this landscape is to understand de Sitter
universe or tunneling processes in the landscape \cite{landscape}.
Our motivation is an attempt to solve some of these questions
within the framework of classical theory of gravity.

The dynamics of the boundary wall of a spherically shaped false
vacuum bubble surrounded by true vacuum regions was originally
studied in Refs.\ \cite{sato01} at the final stage of the true
vacuum bubble nucleation in old inflation \cite{guth01}, and was
studied systematically in Refs.\ \cite{blau01, blau02} as an
attempt to create a universe in the laboratory by quantum
tunneling. They considered the case of interior de Sitter
spacetime and exterior Schwarzschild spacetime divided by
thin-wall (or domain wall). In Ref.\ \cite{chul01} the dynamics of
matter distribution that may contaminate a false vacuum bubble was
considered. The case of interior de Sitter and exterior
Schwarzschild de Sitter spacetime was studied in Ref.\
\cite{aguirre1}, where they examined the instability of false
vacuum bubbles. The case of interior de Sitter spacetime and
exterior Schwarzschild anti-de Sitter spacetime in relation to
AdS/CFT correspondence was considered in Ref.\ \cite{freivogel1}.
The case of charged false vacuum bubble with interior de Sitter
spacetime and exterior Reissner-Nordstr$\rm{\ddot o}$m anti-de
Sitter spacetime with arbitrary dimension was considered in Ref.\
\cite{alberghi1}. The possibility of creation of a universe out of
a monopole in the laboratory was investigated in Ref.\
\cite{sakai1}, where they have considered the classical and
quantum thin-wall dynamics of a magnetic monopole. They have
examined the stability of a spherically symmetric self-gravitating
magnetic monopole in the thin-wall approximation modeling the
interior false vacuum as de Sitter spacetime and the exterior as
the Reissner-Nordstr$\rm{\ddot o}$m spacetime as in Ref.\
\cite{arreaga1}. Recently the classification scheme for possible
evolution of a vacuum wall in the Schwarzschild-de Sitter geometry
was constructed \cite{chernov1}. In addition, there have been
studies on the attempts to create a universe in the laboratory
\cite{borde1}.

As for the false vacuum bubble formation, Lee and Weinberg
\cite{lee01} have shown that gravitational effects make it
possible for bubbles of a higher-energy false vacuum to nucleate
if the vacuum energies are greater than zero. The oscillating
bounce solutions, another type of Euclidean solutions, have been
studied in Refs.\ \cite{banks, hackworth, lavrel}. On the other
hand, Kim {\it et al}.\ \cite{kim01} have shown that there exists
another decay channel which is described by the false vacuum
region of the global monopole formed at the center of a bubble in
the high temperature limit. The Hawking-Moss transition
\cite{haw01}, as another way of vacuum decay, describes the scalar
field jumping simultaneously at the top of the potential barrier.
Recently this process has been interpreted in terms of a thermal
transition \cite{hackworth, linwein01}. It has been shown that the
false vacuum bubble can be nucleated within the true vacuum
background due to a nonminimally coupled scalar field or other
similar coupling terms \cite{wlee01}. The quantum nucleation of
the vacuum bubble was also studied \cite{fisch01}. In Ref.\
\cite{anso01}, they found an analytic expression for the
tunnelling amplitude and studied the tunnelling between arbitrary
(anti-)de Sitter spacetimes in arbitrary spacetime dimensions. The
interesting models for bubble collisions in the very early
universe are also discussed \cite{zcwu}.

In this paper we use the metric junction conditions, which were
developed in Ref.\ \cite{israel01}, to analyze the dynamics of a
false vacuum bubble in the theory with nonminimal coupling. There
are two types of boundary related to this formalism. One is a
``boundary surface'' which is a surface that has the stress-energy
tensor $S_{\mu\nu}=0$ \cite{oppen01}. The process of star
collapsing is a well-known example involving a boundary surface
\cite{oppen02}. The other is a ``surface layer'' which is a thin
layer of matter where $S_{\mu\nu} \neq 0$ \cite{hsato01}. In this
case $S_{\mu\nu}$ is related to the discontinuity of the extrinsic
curvature of the surface. In this framework we classify the
cosmological behaviors from the viewpoint of an observer on the
domain wall and find a solution with multiple accelerations in
five-dimension in the Einstein theory of gravity \cite{nam01}. In
the context of the brane cosmology, after Randall and Sundrum's
interesting proposal \cite{ran01}, the junction conditions have
become one of methods describing the inflationary cosmology on the
brane \cite{kraus01}.

Our approach to obtain junction conditions is based on the method
of variational principle by Chamblin and Reall \cite{cham01}. C.
Barcelo and M. Visser have obtained the generalized junction
conditions using a different approach \cite{barcelo01}.

The plan of this paper is as follows. In Sec.\ 2 we present the
formalism for the junction conditions in the Einstein theory of
gravity with a nonminimally coupled scalar field. In Sec.\ 3 we
study the dynamics of false vacuum bubbles using the junction
conditions. In Sec.\ 4 the bubble wall trajectories in the
exterior bulk spacetime are analyzed according to the mass of the
false vacuum bubble. Finally, we summarize and discuss our results
in Sec.\ 5.

\section{The junction conditions with a nonminimally coupled scalar field  \label{sec2}}

In this section we consider a thin-wall as a surface layer
partitioning bulk spacetime into two distinct four-dimensional
manifolds ${\mathcal M^+}$ and ${\mathcal M^-}$ with boundaries
$\Sigma^+$ and $\Sigma^-$, respectively. To obtain the single
glued manifold ${\mathcal M}={\mathcal M^+} \bigcup {\mathcal
M^-}$ we demand that the boundaries are identified as follows:
\begin{equation}
\Sigma^+ = \Sigma^- = \Sigma ,
\end{equation}
where the thin-wall boundary $\Sigma$ is a timelike hypersurface
with unit normal $n^{\lambda}$.

Let us consider the action
\begin{equation}
S= \int_{\mathcal M} \sqrt{-g} d^4 x \left[
\frac{R}{2\kappa}(1-\xi\kappa\Phi^2)
-\frac{1}{2}{\nabla^\alpha}\Phi {\nabla_\alpha}\Phi
-U(\Phi)\right] + \oint_{\Sigma} \sqrt{-h} d^3 x
\frac{K}{\kappa}(1-\xi\kappa\Phi^2) + S_{tw},
\end{equation}
where $S_{tw}$ is a Nambu-Goto type action on the wall given by $-
\oint_{\Sigma}\sqrt{-h} d^3 x \hat{U}(\Phi)$, $\kappa \equiv 8\pi
G$, and $g\equiv det g_{\mu\nu}$. The second term on the
right-hand side of the above equation is the boundary term
\cite{ygh} with a nonminimally coupled scalar field. $U(\Phi)$ is
the scalar field potential, $R$ denotes the Ricci curvature of
spacetime in ${\mathcal M}$, $K$ is the trace of the extrinsic
curvature of $\Sigma$,  the term $-\xi R\Phi^2 /2$ describes the
nonminimal coupling of the field $\Phi$ to the Ricci curvature and
$\xi$ is a dimensionless coupling constant.

We now vary the action to obtain Israel junction conditions. The
case of minimal coupling has been considered in Ref.\
\cite{cham01}.

Varying the nonminimal coupling term in the bulk $\mathcal M$ we
get
\begin{eqnarray}
\int_{\mathcal M} d^4 x \delta[\sqrt{-g}\xi R\Phi^2 ] & =&
\oint_{\Sigma} \sqrt{-h} d^3 x \xi n^{\lambda} (\Phi^2 g^{\mu\nu}
\nabla_{\mu}\delta g_{\lambda\nu}-\Phi^2
g^{\mu\nu}\nabla_{\lambda}\delta g_{\mu\nu})
\nonumber\\
&+& 2\oint_{\Sigma} \sqrt{-h} d^3 x \xi n_{\lambda}\Phi
[(\nabla^{\lambda}\Phi)g^{\mu\nu}-(\nabla^{\mu}\Phi)g^{\lambda\nu}]\delta
g_{\mu\nu} \nonumber\\
&+& \int_{\mathcal M}\sqrt{-g} d^4 x \xi \Phi^2
(R_{\mu\nu}-\frac{1}{2}g_{\mu\nu}R)\delta g^{\mu\nu} \nonumber\\
&+& 2 \int_{\mathcal M}\sqrt{-g} d^4 x \xi\Phi R \delta\Phi,
\end{eqnarray}
and varying the scalar field action we get
\begin{eqnarray}
\int_{\mathcal M} d^4 x
\delta[\sqrt{-g}g^{\alpha\beta}\nabla_{\alpha}\Phi
\nabla_{\beta}\Phi] &=& 2\oint_{\Sigma} \sqrt{-h} d^3 x
n^{\lambda}(\nabla_{\lambda}\Phi)\delta\Phi -2 \int_{\mathcal
M}\sqrt{-g} d^4 x \nabla^{\lambda}\nabla_{\lambda}\Phi \delta\Phi \nonumber \\
&+& \int_{\mathcal M}\sqrt{-g} d^4 x
(\nabla_{\mu}\Phi\nabla_{\nu}\Phi-\frac{1}{2}\nabla^{\alpha}\Phi\nabla_{\lambda}\Phi
g_{\mu\nu})\delta g^{\mu\nu}.
\end{eqnarray}
The variation of the boundary term with a nonminimal coupling
gives
\begin{eqnarray}
\oint_{\Sigma} \sqrt{-h} d^3 x K(1-\xi\kappa\Phi^2) & = &
\oint_{\Sigma} \sqrt{-h} d^3 x
(1-\xi\kappa\Phi^2)\left[\frac{K}{2}h^{\mu\nu}\delta
g_{\mu\nu}-K^{\alpha\lambda}\delta
g_{\alpha\lambda}-h^{\mu\nu}n^{\lambda}\nabla_{\mu}\delta
g_{\lambda\nu}  \right. \nonumber \\
  \;\;\;\; & +&\left. \frac{1}{2}h^{\mu\nu}n^{\lambda}\nabla_{\lambda}\delta
g_{\mu\nu}+ \frac{K}{2}n^{\mu}n^{\nu}\delta g_{\mu\nu} \right] -
2\oint_{\Sigma} \sqrt{-h} d^3 x K \xi k\Phi \delta\Phi,
\end{eqnarray}
and the variation of the wall action gives
\begin{equation}
\oint_{\Sigma} d^3 x \delta (\sqrt{-h} \hat{U}) = \oint_{\Sigma}
\sqrt{-h} d^3 x h^{\mu\nu}\frac{\hat{U}}{2} \delta g_{\mu\nu} +
\oint_{\Sigma} \sqrt{-h} d^3 x \frac{\partial
\hat{U}}{\partial\Phi}\delta\Phi.
\end{equation}

The bulk Einstein equations are
\begin{equation}
R_{\mu\nu} - \frac{1}{2}g_{\mu\nu}R=\kappa T_{\mu\nu},
\end{equation}
where $R_{\mu\nu}$ is the Ricci tensor and $T_{\mu\nu}$ is the
matter energy momentum tensor,
\begin{eqnarray}
&& T_{\mu\nu} = \frac{1}{1-\xi\Phi^2 \kappa} \left[
\nabla_{\mu}\Phi
\nabla_{\nu}\Phi-g_{\mu\nu}\left(\frac{1}{2}\nabla^{\alpha}\Phi
\nabla_{\alpha}\Phi +U(\Phi)\right)  \right. \nonumber \\
&& \;\;\;\;\;\; + \left. \xi(g_{\mu\nu} \nabla^{\alpha}
\nabla_{\alpha}\Phi^2-\nabla_{\mu} \nabla_{\nu}\Phi^2 )\right].
\end{eqnarray}
The corresponding scalar field equation on the bulk is written by
\begin{equation}
\frac{1}{\sqrt{-g}} \partial_{\mu} [\sqrt{-g}
g^{\mu\nu}\partial_{\nu}\Phi] - \xi R\Phi - \frac{d U}{d\Phi} =0,
\end{equation}
with a boundary condition at the thin-wall
\begin{equation}
n^{\lambda} (\nabla_{\lambda}\Phi) + 2K\xi\kappa\Phi = -
\frac{d\hat{U}}{d\Phi}. \label{sbcat}
\end{equation}
Here we adopt the notations and sign conventions of Misner,
Thorne, and Wheeler \cite{misner}.

The modified Lanczos equation due to a nonminimally coupled scalar
field is given by
\begin{equation}
(1-\xi\kappa\Phi^2_{\pm})([K_{\mu\nu}]-[K]h_{\mu\nu})- 2\xi\kappa
\Phi n^{\lambda} (\nabla_{\lambda}\Phi) h_{\mu\nu} = \kappa
\hat{U} h_{\mu\nu} \label{sjcat}
\end{equation}
where
\begin{equation}
[K] \equiv \lim_{\epsilon \rightarrow 0} K^{+}(\eta = \bar\eta +
\epsilon) - K^{-}(\eta = \bar\eta - \epsilon).
\end{equation}

The sign arises because we have chosen the convention that
$n^{\lambda}$ points towards the region of increasing $\eta$. The
$\bar{\eta}$ is the location of the hypersurface. The signs (+)
and (-) represent exterior and interior spacetime, respectively.

After Eq.\ (\ref{sbcat}) is substituted in Eq.\ (\ref{sjcat}) the
junction conditions become
\begin{equation}
(1-\xi \kappa \Phi^2_+)K^+_{\mu\nu} - (1-\xi \kappa
\Phi^2_-)K^-_{\mu\nu}= - \frac{\kappa}{2} \hat{U} h_{\mu\nu} - \xi
\kappa \left(\Phi_+ \frac{d\Phi_+}{d\eta} -
\Phi_-\frac{d\Phi_-}{d\eta} \right)h_{\mu\nu}. \label{jebwn}
\end{equation}
Actually, the second term on the right-hand side of Eq.\
(\ref{jebwn}) vanishes because $\frac{d\Phi_+}{d\eta}$ and
$\frac{d\Phi_-}{d\eta}$ vanish in the exterier and in the interior
spacetime of the wall, respectively.

In order to find the gravitational field and the motion of a wall
we must first find two sets, both inside and outside of the wall,
of solutions of the bulk Einstein equation and scalar field
equation. So if the bulk solutions are given we only need to match
the junction conditions to determine the motion of the wall. In
the next section, we will only consider the junction equations
because bulk solutions are easily given; the bulk solution of
$M=0$ is already known in Ref.\ \cite{wlee01}, while the case for
$M \neq 0$ corresponds to Schwarzschild solution because of
Birkhoff's theorem \cite{birkhoff}. For the case of $M < 0$, the
mass of a false vacuum becomes effectively negative, which is
possible due to nonminimal coupling.

\section{Dynamics of a false vacuum bubble \label{sec3}}

For applications of the modified junction equations on the false
vacuum bubble, the bulk spacetime geometry for the inside$(-)$ and
outside$(+)$ of the wall have a spherically symmetric spacetime
\begin{equation}
ds^2 = -H_{\pm}(R) dT^2 + \frac{dR^2}{H_{\pm}(R)} + R^2 d\Omega^2
, \label{dfvb01}
\end{equation}
where
\begin{equation}
H_-=1-A_{-}R^2  \;\;\; {\rm and} \;\;\; H_+=1-A_{+}R^2 -
\frac{2GM}{R},
\end{equation}
and $M$ is the mass or the total energy of a false vacuum bubble.
The constant $A$ is related to a cosmological constant; $A= +
\frac{\Lambda}{3}=+ \frac{8\pi G}{3}\rho$ for de Sitter spacetime
, $A=0$ for Minkowski spacetime, and $A= -\frac{\Lambda}{3}=-
\frac{8\pi G}{3}\rho$ for anti-de Sitter spacetime. Since we
consider a false vacuum bubble, $\rho_{-} > \rho_{+}$.

We take the energy-momentum tensor $T^{\mu\nu}$ as the form
\begin{equation}
T^{\mu\nu} = S^{\mu\nu} \delta(\eta) + (\rm{regular \; terms}),
\end{equation}
where $S^{\mu\nu}=-\sigma h^{\mu\nu} (x^i, \eta=\bar\eta)$ and
$\sigma$ is the positive surface energy density, or surface
tension, of the wall without nonminimal coupling. For bubble walls
$\sigma$ is a constant having the same value at all events on the
timelike surface \cite{blau01, ipser}. Note that the stress-energy
tensor of the surface $S_{\mu\nu}$ can be defined as the integral
over the thickness, $\epsilon$, of the surface $\Sigma$ in the
limit as $\epsilon$ goes to zero
\begin{equation}
S_{\mu\nu} = \lim_{\epsilon \rightarrow 0} \int^{\bar\eta +
\epsilon}_{\bar\eta - \epsilon} T_{\mu\nu} d\eta .
\end{equation}
Then, ${\hat U} = \sigma$ because the internal structure of the
wall is neglected in thin-wall limit and $h^{\eta\eta}=h^{\eta
i}=0$.

To keep the analysis as simple as possible, we take the position
of a false vacuum in the potential as zero, that is $\Phi_-=0$(see
Ref.\ \cite{wlee01}). In this case Eq.\ (\ref{jebwn}) becomes
\begin{equation}
(1-\xi \kappa \Phi^2_+)K^{+ i}_{j} - K^{- i}_{j}= - 4\pi G \sigma
\delta^i_j .
\end{equation}

By spherical symmetry, the extrinsic curvature has only two
components, $K^{\theta}_{\theta} \equiv K^{\phi}_{\phi}$ and
$K^{\tau}_{\tau}$. The junction equation is related to
$K^{\theta}_{\theta}$ and the covariant acceleration in the normal
direction is related to $K^{\tau}_{\tau}$.

We introduce the Gaussian normal coordinate system near the wall
\begin{equation}
dS^2 = - d\tau^2 + d\eta^2 + {\bar r}^2(\tau, \eta) d\Omega^2,
\end{equation}
where $g_{\tau\tau}=-1$ and ${\bar r}(\tau, \bar{\eta})=r(\tau)$.
It must agree with the coordinate $R$ of the interior and exterior
coordinate systems. The angle variables can be taken to be
invariant in all regions. In this coordinate system the induced
metric on the wall is given by
\begin{equation}
dS^2_{\Sigma} = - d\tau^2 + r^2(\tau) d\Omega^2,
\end{equation}
where $\tau$ is the proper time measured by an observer at rest
with respect to the wall and $r(\tau)$ is the proper radius of
$\Sigma$. The following relation is satisfied
\begin{equation}
d\tau^2 = H_{\pm}(R)dT^2 - \frac{dR^2}{H_{\pm}(R)}.
\end{equation}

In these treatments, the condition becomes
\begin{equation}
\sqrt{\dot{r}^2 + H_-} - (1-\xi \kappa \Phi^2_+)\sqrt{\dot{r}^2 +
H_+} = \frac{1}{2}\kappa \sigma r. \label{ntdnc0}
\end{equation}
or more generally
\begin{equation}
\epsilon_- \sqrt{\dot{r}^2 + H_-} - \epsilon_+ \sqrt{\dot{r}^2 +
H_+} = \frac{1}{2}\kappa r(\sigma - \xi \bar\sigma ),
\label{ntdnc}
\end{equation}
where $\bar\sigma = \frac{2 c}{r}\sqrt{\dot{r}^2 + H_+}$.
Hereafter $c$ denotes $\Phi^2_+$. We are using the dot notation to
refer to derivatives with respect to $\tau$. $\epsilon_{\pm}$ are
$+1$ if the outward normal to the wall is pointing towards
increasing $r$ and $-1$ if towards decreasing $r$ \cite{alberghi1,
boul1}. There are parameter regions where both $\epsilon_-$ and
$\epsilon_+$ are positive in all ranges of $r$. This situation is
similar to the case of the evolution of a true vacuum bubble. In
earlier works, a sign change of $\epsilon_{\pm}$ was needed to
cover all ranges of $r$ for the solution. This is because the
positive signs for $\epsilon_{\pm}$ covered only partial ranges of
$r$; thus the interesting unbound solutions were excluded. To
obtain an expanding false vacuum bubble, the mass of the bubble
should be greater than some critical value. These bubbles start
from an initial singularity. The second term on the right-hand
side of Eq.\ (\ref{ntdnc}) can be interpreted as the negative
tension of the wall due to a nonminimal coupling term.

After squaring twice, the Eq.\ (\ref{ntdnc0}) can be written in
the form
\begin{equation}
\frac{1}{2} {\dot r}^2 + V_{eff}(r) = 0, \label{emp01}
\end{equation}
where the effective potential is
\begin{equation}
V_{eff}(r)= \frac{T+\sqrt{T^2 - PQ}}{2P}, \label{ep03}
\end{equation}
with
\begin{eqnarray}
T&=&[1-(1-\xi \kappa c)^2]^2 + \frac{2GM}{r}[1-(1-\xi \kappa
c)^2](1-\xi \kappa c)^2  \nonumber \\
& + & \{[1-(1-\xi \kappa c)^2][(1-\xi \kappa c)^2 A_+ - A_- +
\frac{1}{4} \kappa^2\sigma^2] - \frac{1}{2}\kappa^2\sigma^2 \}r^2 ,  \nonumber \\
Q &=&\{[(1-\xi \kappa c)^2 A_+ - A_- + \frac{1}{4}\kappa^2
\sigma^2 ]^2 +
 A_-\kappa^2\sigma^2 \} r^4   \nonumber \\
&+& 2 \{[1-(1-\xi \kappa c)^2][(1-\xi \kappa c)^2 A_+ - A_- +
\frac{1}{4} \kappa^2\sigma^2] - \frac{1}{2}\kappa^2\sigma^2 \}r^2  \nonumber \\
&+& [(1-\xi \kappa c)^2 A_+ - A_- + \frac{1}{4}\kappa^2
\sigma^2](1-\xi \kappa c)^2 4GM r + [1-(1-\xi \kappa c)^2]^2 \nonumber \\
&+& [1-(1-\xi \kappa c)^2](1-\xi \kappa c)^2 \frac{4GM}{r} +
(1-\xi \kappa c)^4
\frac{4G^2M^2}{r^2},  \nonumber \\
P&=&[1-(1-\xi \kappa c)^2]^2.
\end{eqnarray}
Eq.\ (\ref{emp01}) formally coincides with the equation describing
one-dimensional motion of a unit-mass particle moving in the
corresponding potential $V_{eff}$ with zero total energy. The
properties of the trajectory of the wall can be read off directly
from the shape of $V_{eff}$. In the next section we will discuss
the details of trajectories of the wall.

\section{The bubble wall trajectories \label{sec4}}

In this section we consider the bubble wall trajectories according
to the mass of a false vacuum bubble in the exterior bulk
spacetime. The modified junction equations with nonminimal
coupling determine the trajectories. From the shape of $V_{eff}$
we can obtain the behavior of solutions without solving the
equations exactly. We consider only bubble solutions without black
holes. In other words, we consider the size of a false vacuum
bubble larger than the black hole horizon.

\subsection {The case of $M=0$}

\begin{figure}[t]
\begin{center}
\includegraphics[width=2.9in]{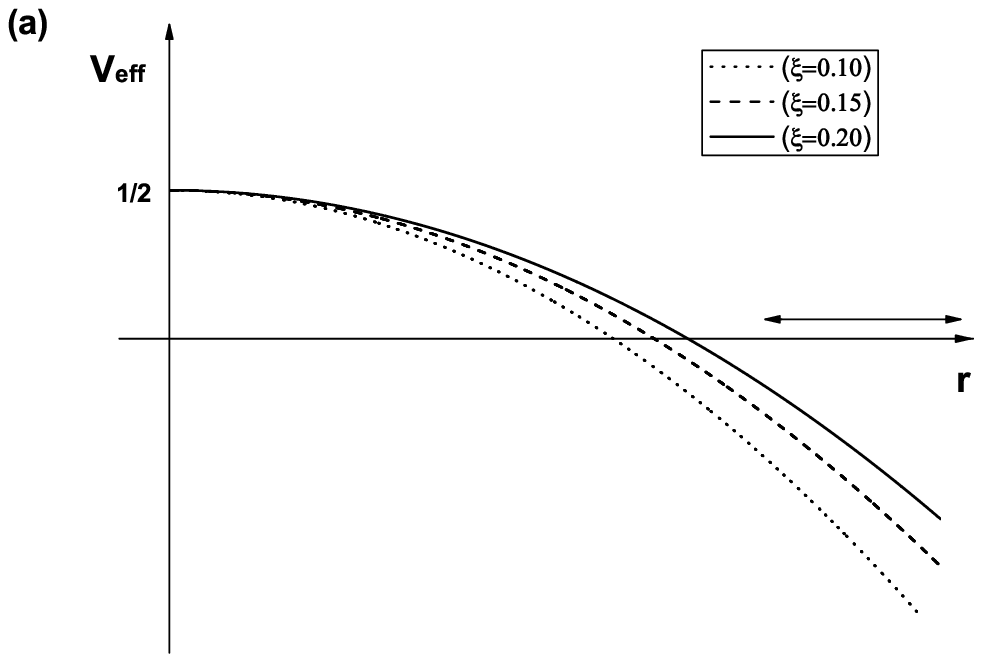}
\includegraphics[width=2.9in]{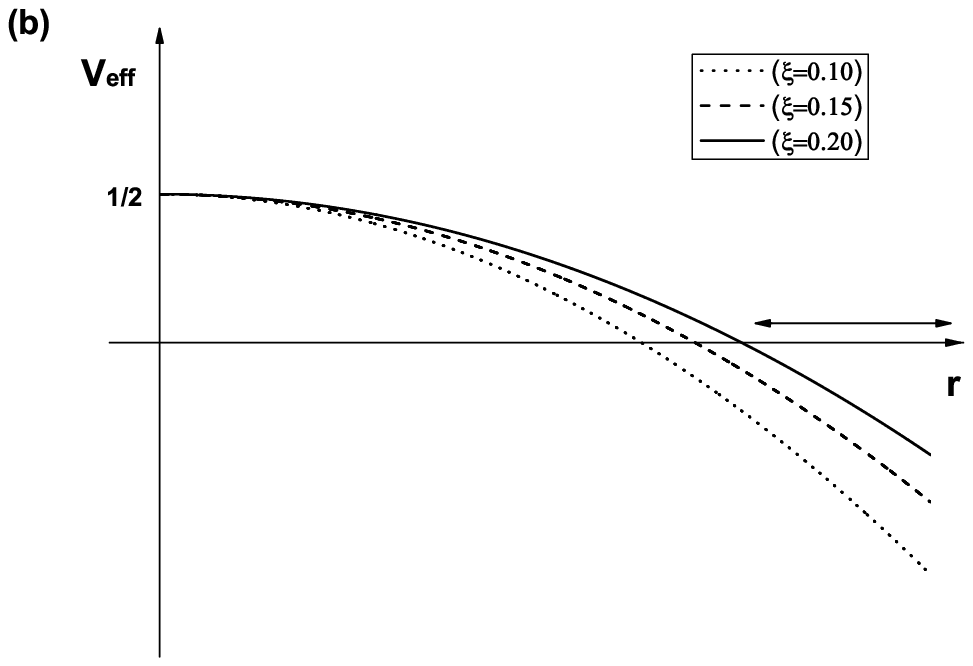}\\
\includegraphics[width=2.9in]{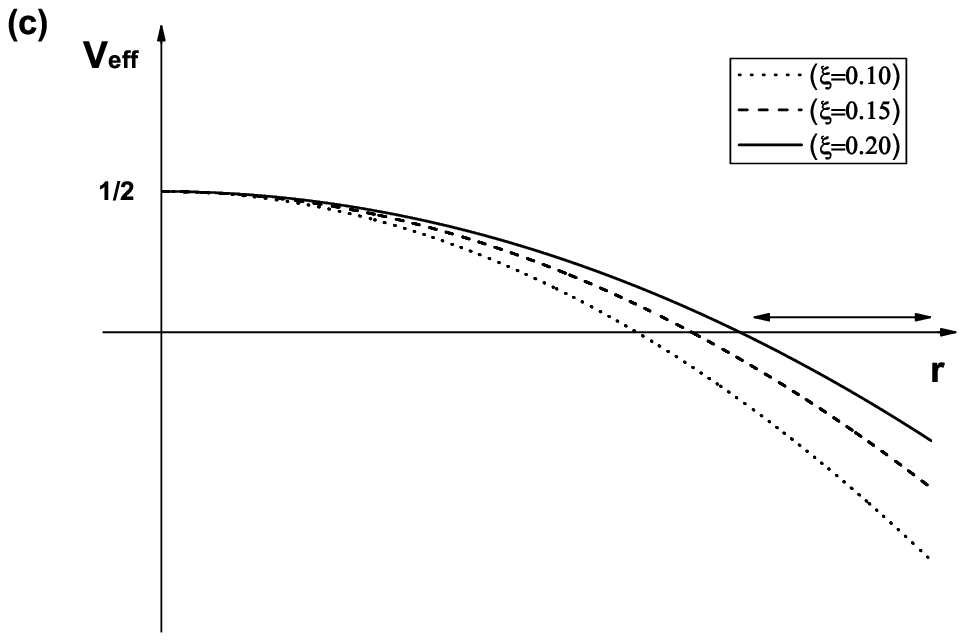}
\includegraphics[width=2.9in]{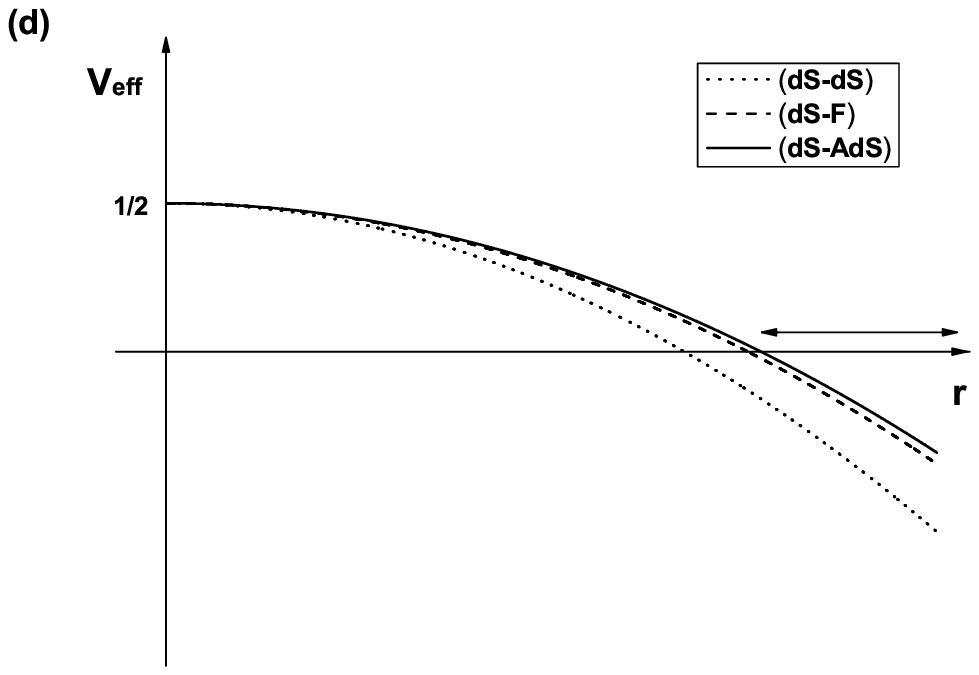}
\end{center}
\caption{\footnotesize{The effective potential $V_{eff}$ for
various $\xi$ in the case of $M=0$. The time evolution of the wall
can be interpreted as the motion of a particle
 coming from infinity, reflecting at the barrier, and then
going back to infinity. The three curves are (i) dotted curve:
$\xi=0.10$; (ii) dashed curve: $\xi=0.15$; (iii) solid curve:
$\xi=0.20$ in (a), (b) and (c). Figure (d) indicates the potential
with different background at $\xi=0.20$. There exist only
"unbound" solutions. }} \label{fig:fig1}
\end{figure}

\begin{figure}[t]
\begin{center}
\includegraphics[width=2.9in]{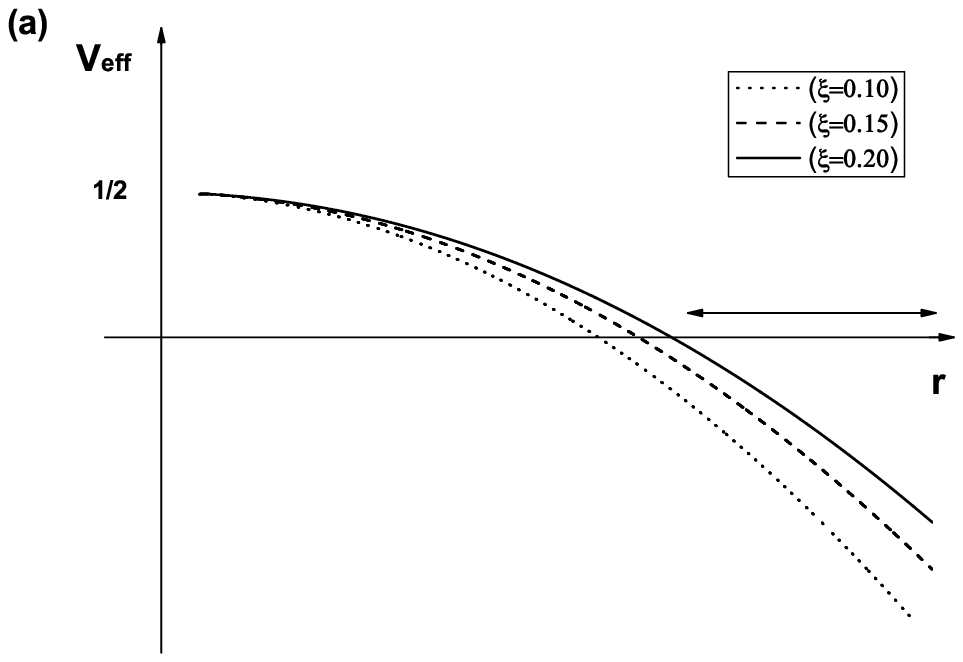}
\includegraphics[width=2.9in]{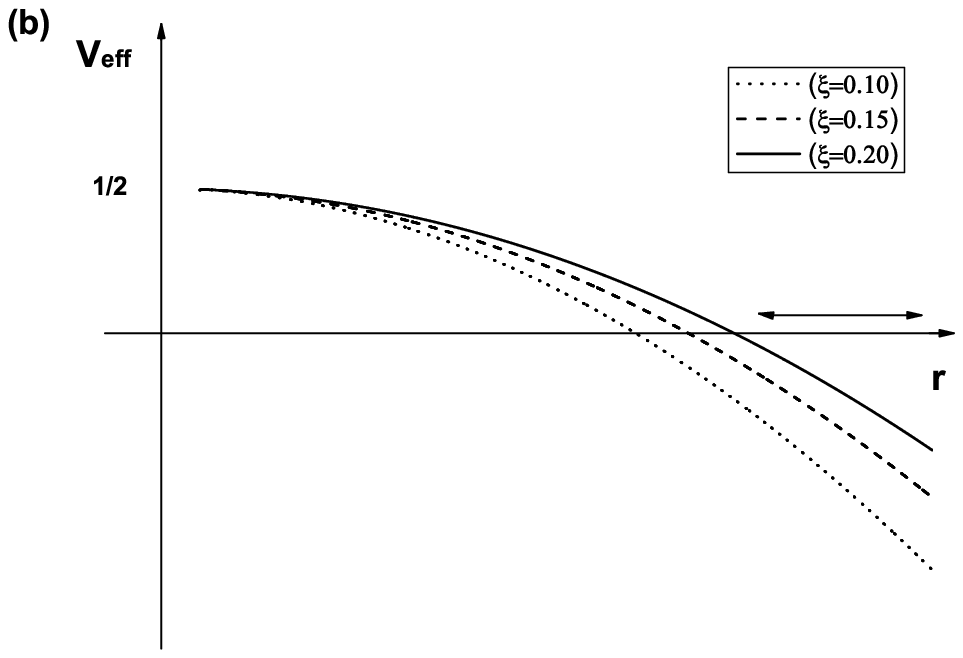}\\
\includegraphics[width=2.9in]{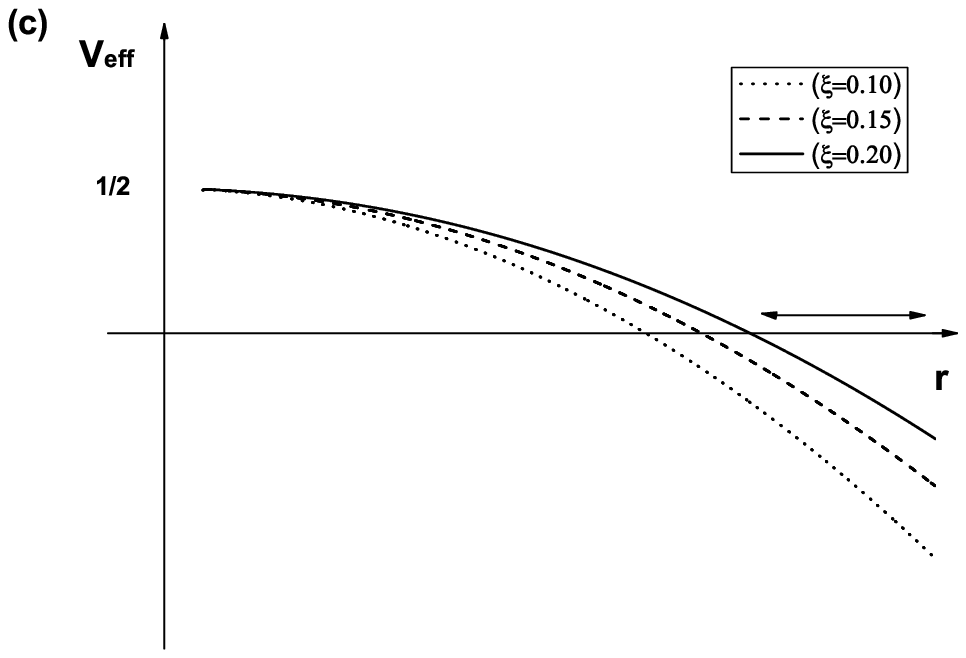}
\includegraphics[width=2.9in]{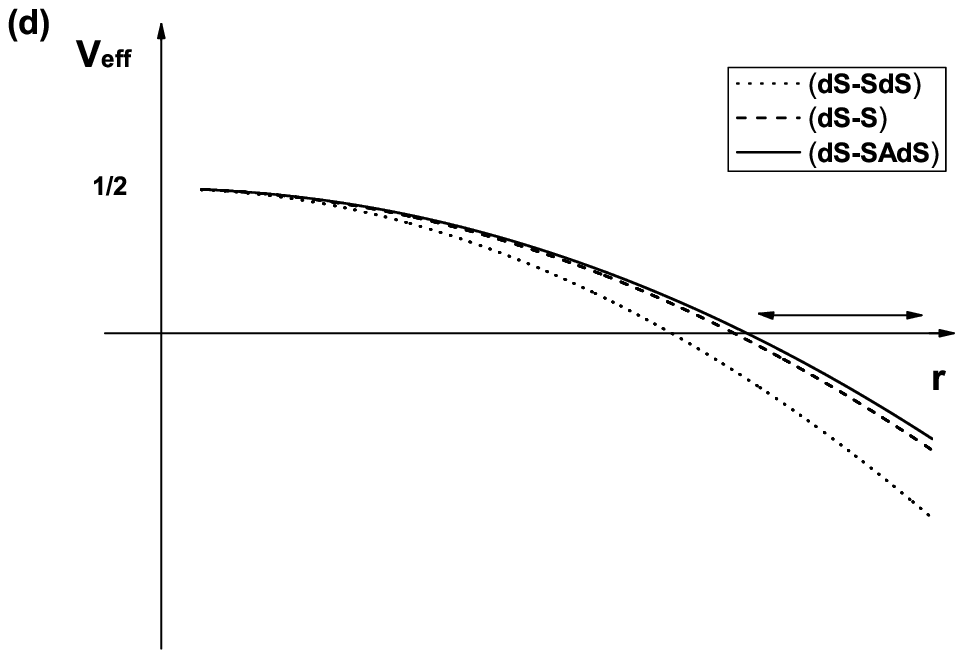}
\end{center}
\caption{\footnotesize{The effective potential $V_{eff}$ for
various $\xi$ in the case of $M > 0$. The three curves are (i)
dotted curve: $\xi=0.10$; (ii) dashed curve: $\xi=0.15$; (iii)
solid curve: $\xi=0.20$ in (a), (b) and (c). There exist only
"unbound" solutions.}} \label{fig:fig2}
\end{figure}

This case is related to the results in Ref.\ \cite{wlee01}. If a
false vacuum bubble can be nucleated within the true vacuum
background without changing the exterior spacetime, the surface
density becomes negative and/or junction conditions itself need to
be modified. Our results show how the surface tension as well as
junction conditions are modified by nonminimal coupling. From now
on we scale the dimension of a vacuum energy density to $M^4$,
that of $\sigma$ to $M^3$, that of $G$ to $M^{-2}$, that of $r$ to
$M^{-1}$, and that of $c$ to $M^2$, to make all terms in Eq.\
(\ref{emp01}) dimensionless. For the sake of simplicity we take
$\kappa c = 0.1$, $\kappa\sigma=0.033$, and $A_- - A_+ = 0.0025$.
In this case the effective potential function is $V_{eff}(0)=
\frac{1}{2}$ at $r=0$ and there exist only "unbound" solutions. We
consider three particular cases: (case 1) the interior as well as
the exterior spacetime is de Sitter; (case 2) the interior false
vacuum as de Sitter and the exterior as flat Minkowski spacetime;
(case 3) the interior false vacuum as de Sitter and the exterior
as anti-de Sitter spacetime. The shapes of the effective potential
as a function of $r$ are shown in Fig.\ \ref{fig:fig1}. These
figures indicate only unbound trajectories are possible. That is,
the bound and monotonic trajectories do not appear as classical
solutions.

We see that the allowed minimum size of a false vacuum bubble is
diminished as $\xi$ is decreased. So if the radius of a nucleated
false vacuum bubble is greater than the allowed minimum size then
the false vacuum bubble can expand within the true vacuum
background. These expanding bubbles have no initial singularity,
as we can see from the Fig.\ \ref{fig:fig1}. So it is possible to
create a universe by an expanding false vacuum bubble nucleated by
a proper mechanism.

\subsection {The case of $M > 0$}

\begin{figure}[t]
\begin{center}
\includegraphics[width=2.9in]{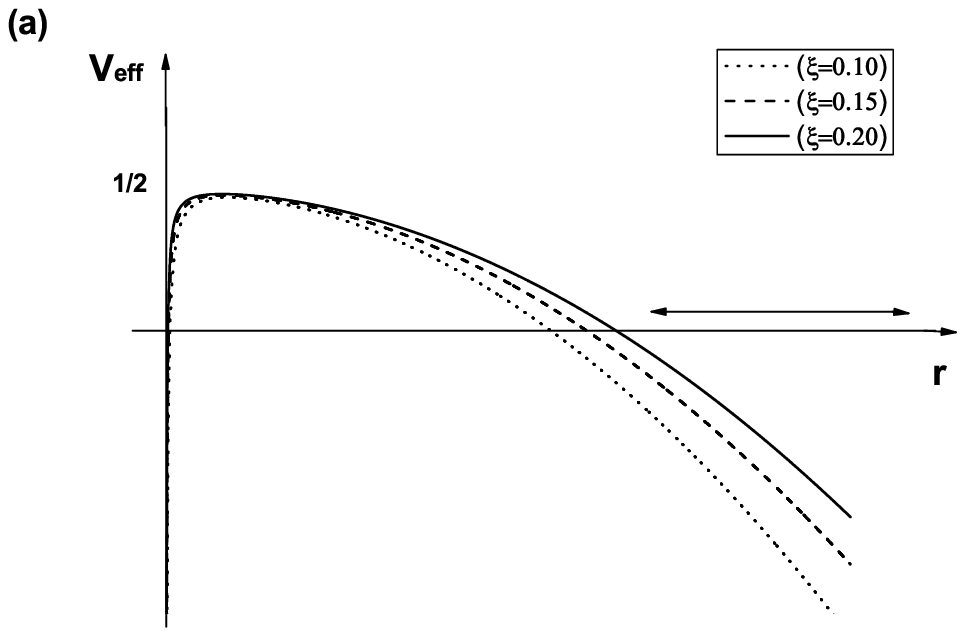}
\includegraphics[width=2.9in]{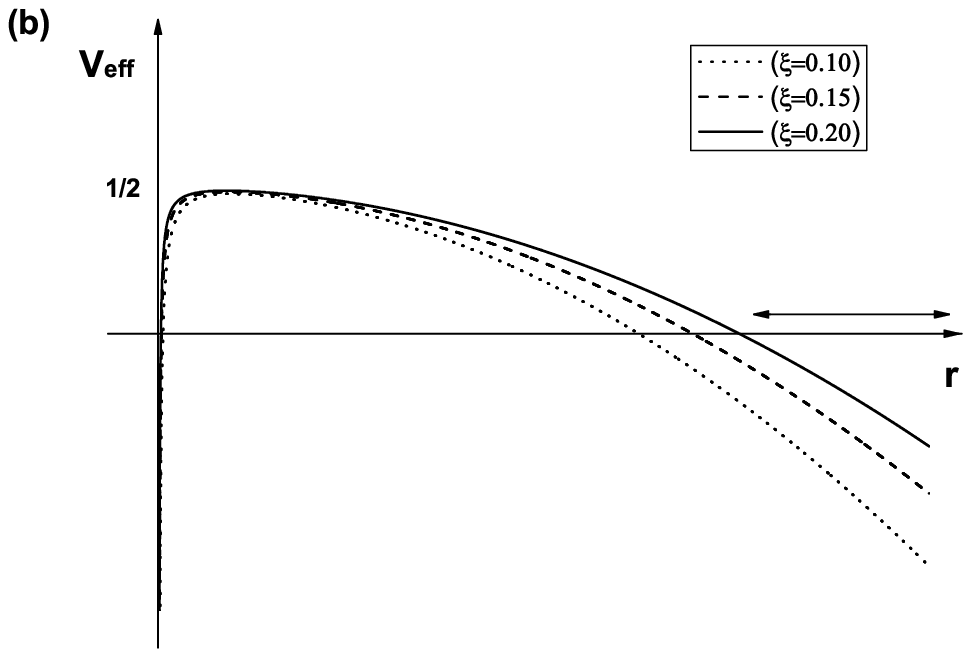}\\
\includegraphics[width=2.9in]{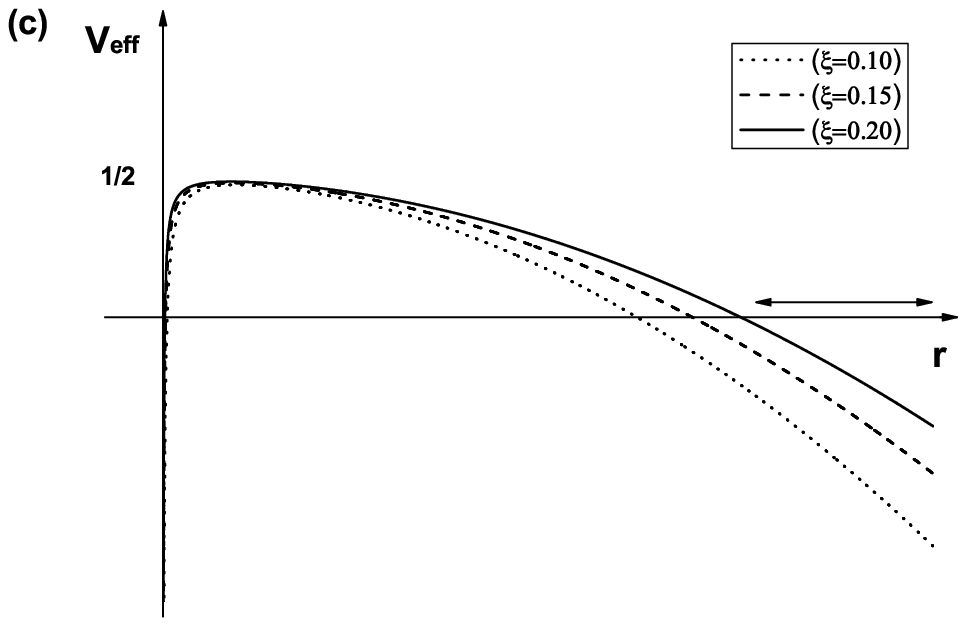}
\includegraphics[width=2.9in]{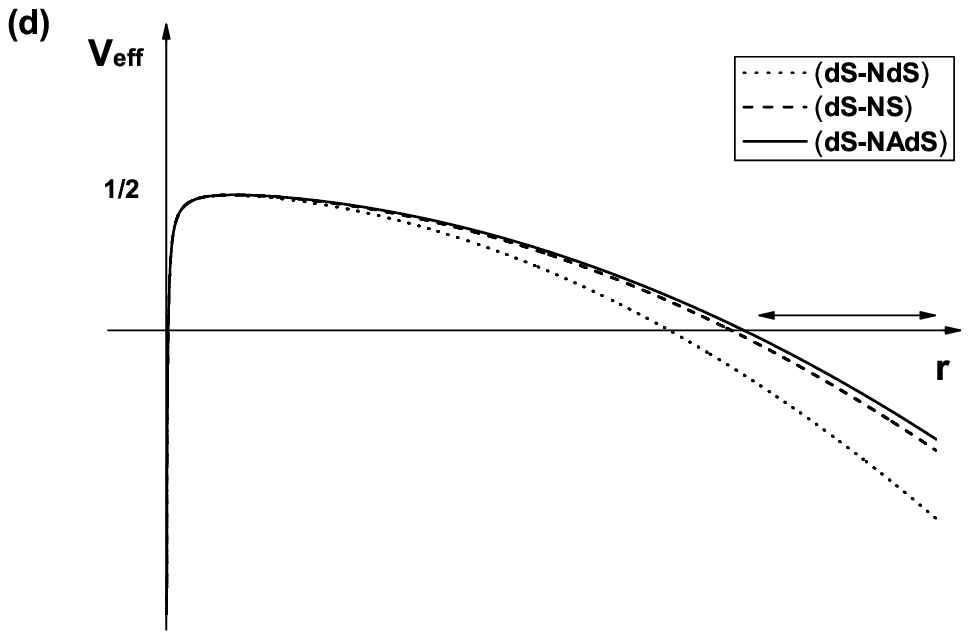}
\end{center}
\caption{\footnotesize{The effective potential $V_{eff}$ for
various $\xi$ in the case of $M < 0$. The three curves are (i)
dotted curve: $\xi=0.10$; (ii) dashed curve: $\xi=0.15$; (iii)
solid curve: $\xi=0.20$ in (a), (b) and (c). There exist "unbound"
solutions as well as "bound" solutions. }} \label{fig:fig3}
\end{figure}

In this case the mass of the false vacuum bubble is taken as a
constant parameter. The portion of the potential is inherently
restricted since the allowed region of $r$ from Eq.\ (\ref{ep03})
is given by
\begin{equation}
r > \left( \frac{2GM[1-(1-\xi \kappa c)^2](1-\xi \kappa
c)^2}{\frac{1}{4}\kappa^2\sigma^2(1-\xi \kappa c)^2-[1-(1-\xi
\kappa c)^2][(1-\xi \kappa c)^2A_+ -A_-]} \right)^{1/3}.
\label{eq27}
\end{equation}

In the case $M > 0$, we consider three particular cases: (case 1)
the interior spacetime as de Sitter and the exterior as
Schwarzschild de Sitter; (case 2) the interior as de Sitter and
the exterior as Schwarzschild spacetime; (case 3) the interior as
de Sitter and the exterior as Schwarzschild anti-de Sitter
spacetime. For these cases massive bubbles can be formed in the
early universe. These cases have been studied by many authors in
the pure Einstein theory of gravity \cite{blau01, blau02,
aguirre1, freivogel1}. However, there are different features
between their models and ours. One is related to the sign of
$\epsilon_{\pm}$. Unlike previous works unbound solutions are
allowed in our model since there are parameter regions where both
$\epsilon_-$ and $\epsilon_+$ are positive in all ranges of $r$.
Note that there is the restricted region of $r$ as in Eq.\
(\ref{eq27}). The other is related to the junction equation Eq.\
(\ref{emp01}). Our approach is somewhat different from their
works. We consider the case of positive as well as zero mass. It
seems not appropriate that the case of zero mass is applied in
their formalism. We can consider the junction equation regardless
of the mass. The shapes of the effective potential as a function
of $r$ are shown in Fig.\ \ref{fig:fig2}. In these cases, the
false vacuum bubble can also expand within the true vacuum
background.

\subsection {The case of $M < 0$ }

The case of negative mass bubble is allowed in this framework. In
this case we assume the geometry of outside spacetime with
spherical symmetry is similarly written by Eq.\ (\ref{dfvb01}).
These objects are considered in different contexts in Refs.\
\cite{mann2}.

In the case of $M < 0$, we consider three particular cases: (case
1) the interior spacetime as de Sitter and the exterior as
Schwarzschild, with negative mass, de Sitter; (case 2) the
interior as de Sitter and the exterior as Schwarzschild spacetime;
(case 3) the interior as de Sitter and the exterior as
Schwarzschild anti-de Sitter spacetime. Although the case of
negative mass is physically unclear, since it does not satisfy the
positive energy condition, it still gives rise to the solutions of
the Einstein equations. So we proceed to analyze the case of $M <
0$. The shapes of the effective potential as a function of $r$ are
shown in Fig.\ \ref{fig:fig3}. There also exist expanding false
vacuum bubbles.

\begin{figure}[t]
\begin{center}
\includegraphics[width=4.5in]{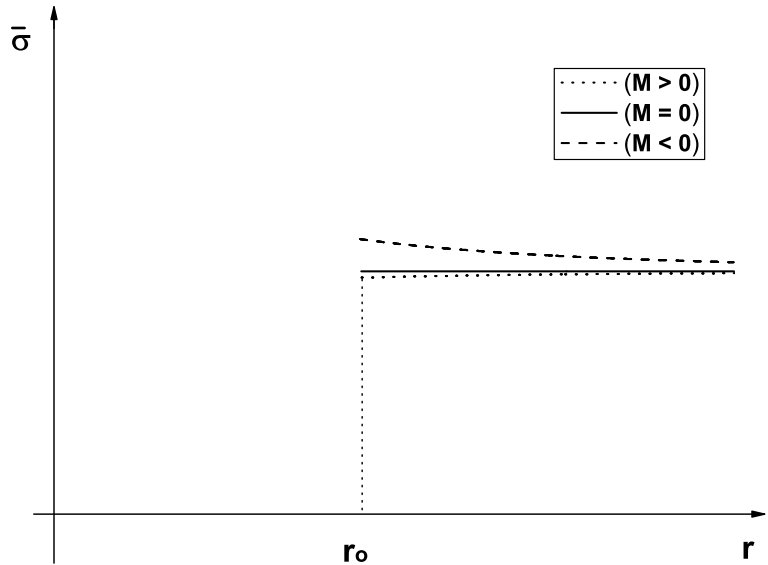}
\end{center}
\caption{\footnotesize{The magnitude of the negative tension of
the wall, $\bar{\sigma}$, due to a nonminimal coupling term in the
spacetime with different mass sign. Here $r_o$ denotes the allowed
minimum size of a false vacuum bubble.}} \label{fig:fig4}
\end{figure}

Here we discuss again the term which can be interpreted as the
negative tension of the wall. This effect is different from
earlier works \cite{blau01, blau02, aguirre1, freivogel1}. In
Fig.\ \ref{fig:fig4}, we see that the magnitude of the negative
tension of the wall, $\bar{\sigma}$, due to a nonminimal coupling
term is a constant only in the case of $M=0$. In other cases, the
magnitudes approach the value in the case of $M=0$ as $r$ is
increased. On the other hand the magnitude is increased as $r$ is
decreased in the case of $M < 0$ and decreased as $r$ is decreased
in the case of $M > 0$. Unlike other cases, there exist bound
solutions in narrow range of $r$ in the case of $M < 0$ although
we are not interested in this. In Fig.\ \ref{fig:fig4} we take
only unbound solutions. $r_o$ denotes the allowed minimum size of
a false vacuum bubble in this framework.

\section{Summary and Discussions}

In this paper we have shown that there can be an expanding false
vacuum bubble within the true vacuum background. We have presented
the formalism for the junction conditions with nonminimal coupling
in section 2.

In section 3, we have studied the dynamics of a false vacuum
bubble using the modified junction conditions. The nonminimal
coupling term can be interpreted as the negative tension of the
wall in the junction conditions. In this treatment the mass of a
false vacuum bubble from outside observer's point of view can be
positive, zero, or negative. The solutions in our model does not
have black holes. In other words, the size of a false vacuum
bubble is larger than the black hole horizon. The mass of a false
vacuum bubble has been treated as a parameter in this work.

In section 4, the bubble wall trajectories in the exterior bulk
spacetime are analyzed according to the mass of false vacuum
bubbles. We have obtained the behavior of solutions in various
cases. In this framework there are parameter regions where both
$\epsilon_-$ and $\epsilon_+$ are positive in all ranges of $r$.
This situation is similar to the case of the evolution of a true
vacuum bubble. In the case of $M=0$, we have considered three
particular cases: (case 1) the interior as well as the exterior
spacetime is de Sitter; (case 2) the interior false vacuum as de
Sitter and the exterior as flat Minkowski spacetime; (case 3) the
interior false vacuum as de Sitter and the exterior as anti-de
Sitter spacetime. In these cases only unbound trajectories are
possible. In the case $M > 0$, we have considered three particular
cases: (case 1) the interior spacetime as de Sitter and the
exterior as Schwarzschild de Sitter; (case 2) the interior as de
Sitter and the exterior as Schwarzschild spacetime; (case 3) the
interior as de Sitter and the exterior as Schwarzschild anti-de
Sitter spacetime. In these cases also only unbound trajectories
are possible. The portion of the potential is inherently
restricted since the allowed region of $r$ is given by Eq.\
(\ref{eq27}). For the case of $M < 0$, there also exist expanding
false vacuum bubbles. These objects are considered in different
contexts in Ref.\ \cite{mann2}.

In earlier works on the expanding false vacuum bubbles
\cite{blau01, blau02, aguirre1, freivogel1}, they have the initial
singularity in the past. In Ref.\ \cite{wlee01}, it was shown that
a false vacuum bubble can be nucleated within the true vacuum
background due to a nonminimally coupled scalar field. In order to
keep the outside geometry invariant, after a false vacuum bubble
is nucleated, the junction conditions need to be modified. In our
model the false vacuum bubbles with minimal coupling can expand
within the true vacuum background with nonminimal coupling. It
will be interesting if this solution can be related to tunnelling
from nothing to de Sitter space \cite{vil03} or related to a kind
of eternal inflation \cite{eternal}. Our model is within a
framework of classical theory of gravity. It will be interesting
if this framework can be embedded in the superstring theory.

\section*{Acknowledgements}

We would like to thank Yun Soo Myung, Yongsung Yoon, Hongsu Kim,
Sunggeun Lee, John J. Oh, Myungseok Yoon and Changheon Oh for
valuable discussions and kind comments. We would like to thank S.
Ansoldi and S. Sasaki for their constructive comments and M.
Sasaki for his hospitality at the YITP-KIAS joint workshop on
String phenomenology and Cosmology, September, 2007. We would like
to thank E. J. Weinberg for helpful discussions. This work was
supported by the Science Research Center Program of the Korea
Science and Engineering Foundation through the Center for Quantum
Spacetime (CQUeST) of Sogang University with grant number R11 -
2005- 021 and by the Sogang University Research Grant in 2007. WL
was supported by the Korea Research Foundation Grant funded by the
Korean Government(MOEHRD)(KRF-2007-355-C00014).


\newpage

\begin{thebibliography}{99}
\bibitem{wheeler} J. A. Wheeler, Ann. Phys. {\bf 2}, 604 (1957).
\bibitem{blau01} S. K. Blau, E. I. Guendelman, and A. H. Guth, Phys. Rev. D {\bf 35},
1747 (1987).
\bibitem{blau02} E. Farhi and A. H. Guth, Phys. Lett. B{\bf 183}, 149
(1987); E. Farhi, A. H. Guth, and J. Guven, Nucl. Phys. B{\bf
339}, 417 (1990).
\bibitem{aguirre1} A. Aguirre and M. C. Johnson, Phys. Rev. D {\bf 72}, 103525
(2005).
\bibitem{freivogel1} B. Freivogel, V. E. Hubeny, A. Maloney, R. Myers, M. Rangamani, and S. Shenker, J. High Energy Phys. 03 (2006) 007.
\bibitem{banks} T. Banks, hep-th/0211160.
\bibitem{entro01} R. Bousso, Phys. Rev. D {\bf 71}, 064024 (2005).
\bibitem{susskind} L. Susskind, hep-th/0302219.
\bibitem{landscape} S. Kachru, R. Kallosh, A. Linde, and S. P.
Trivedi, Phys. Rev. D {\bf 68}, 046005 (2003); J. Garriga and A.
Megevand, Int. J. Theor. Phys. {\bf 43}, 883 (2004); B. Freivogel
and L. Susskind, Phys. Rev. D {\bf 70}, 126007 (2004); J. Garriga,
D. Schwartz-Perlov, A. Vilenkin, and S. Winitzki, J. Cosmol.
Astropart. Phys. 01 (2006) 017; B. Freivogel, M. Kleban, M. R.
Martinez, and L. Susskind, J. High Energy Phys. 03 (2006) 039; S.
W. Hawking and T. Hertog, Phys. Rev. D {\bf 73}, 123527 (2006); V.
Vanchurin and A. Vilenkin, Phys. Rev. D {\bf 74}, 043520 (2006);
A. Ceresole, G. Dall'Agata, A. Giryavets, R. Kallosh, and A.
Linde, Phys. Rev. D {\bf 74}, 086010 (2006); P. Batra and M.
Kleban, hep-th/0612083; E. J. Weinberg, Phys. Rev. Lett. {\bf 98},
251303 (2007); A. Linde, J. Cosmol. Astropart. Phys. 01 (2007)
022; T. Clifton, A. Linde, and N. Sivanandam, J. High Energy Phys.
02 (2007) 024; A. Aguirre, S. Gratton, and M. C. Johnson, Phys.
Rev. D {\bf 75}, 123501 (2007); D. Podolsky and K. Enqvist,
arXiv:0704.0144.
\bibitem{sato01} K. Sato, M. Sasaki, H. Kodama, and K. Maeda, Prog. Theor. Phys. {\bf 65}, 1443
(1981); H. Kodama, M. Sasaki, K. Sato, and K. Maeda, {\it ibid.}
{\bf 66}, 2052 (1981); K. Maeda, K. Sato, M. Sasaki, and H.
Kodama, Phys. Lett. {\bf 108}B, 98 (1982); K. Sato, H. Kodama, M.
Sasaki, and K. Maeda, {\it ibid.} {\bf 108}B, 103 (1982).
\bibitem{guth01} A. H. Guth, Phys. Rev. D {\bf 23}, 347
(1981); K. Sato, Mon. Not. R. Astron. Soc. {\bf 195}, 467 (1981).
\bibitem{chul01} C. H. Lee, Mod. Phys. Lett. A {\bf 7}, 419 (1992).
\bibitem{alberghi1} G. L. Alberghi, D. A. Lowe, and M. Trodden, J. High Energy Phys. 07 (1999) 020.
\bibitem{sakai1} N. Sakai, K.-i. Nakao, H. Ishihara, and M. Kobayashi, Phys. Rev. D {\bf 74}, 024026 (2006).
\bibitem{arreaga1} G. Arreaga, I. Cho, and J. Guven, Phys. Rev. D {\bf 62}, 043520 (2000).
\bibitem{chernov1} S. V. Chernov and V. I. Dokuchaev,
arXiv:0709.0616.
\bibitem{borde1} A. Borde, M. Trodden, and T. Vachaspati, Phys. Rev. D {\bf 59}, 043513
(1999); E. I. Guendelman and J. Portnoy, Class. Quantum Grav. {\bf
16}, 3315 (1999); E. I. Guendelman and J. Portnoy, Mod. Phys.
Lett. A {\bf 16}, 1079 (2001); I. G. Dymnikova, A. Dobosz, M. L.
Fil'chenkev, and A. Gromov, Phys. Lett. B {\bf 506}, 351 (2001);
A. Aguirre and M. C. Johnson, Phys. Rev. D {\bf 73}, 123529
(2006); S. Ansoldi and E. I. Guendelman, gr-qc/0611034.
\bibitem{lee01} K. Lee and E. J. Weinberg, Phys. Rev.
D {\bf 36}, 1088 (1987).
\bibitem{hackworth} J. C. Hackworth and E. J. Weinberg, Phys. Rev. D {\bf 71},
044014 (2005).
\bibitem{lavrel} G. Lavrelashvili, Phys. Rev. D {\bf 73}, 083513
(2006).
\bibitem{kim01} Y. Kim, K. Maeda, and N. Sakai, Nucl. Phys.
B{\bf 481}, 453 (1996); Y. Kim, S. J. Lee, K. Maeda, and N. Sakai,
Phys. Lett. B{\bf 452}, 214 (1999).
\bibitem{haw01} S. W. Hawking and I. G. Moss, Phys. Lett. {\bf 110B}, 35 (1982).
\bibitem{linwein01} A. R. Brown and E. J. Weinberg, Phys. Rev. D {\bf
76}, 064003 (2007).
\bibitem{wlee01} W. Lee, B.-H. Lee, C. H. Lee, and C. Park, Phys.
Rev. D {\bf 74}, 123520 (2006).
\bibitem{fisch01} W. Fischler, D. Morgan, and J. Polchinski, Phys.
Rev. D {\bf 41}, 2638 (1990); A. Aurilia and E. Spallucci, Phys.
Lett. B {\bf 251}, 39 (1990); W. Fischler, D. Morgan, and J.
Polchinski, Phys. Rev. D {\bf 42}, 4042 (1990); A. Aurilia, R.
Balbinot, and E. Spallucci, Phys. Lett. B {\bf 262}, 222 (1991).
\bibitem{anso01} S. Ansoldi and L. Sindoni, Phys.
Rev. D {\bf 76}, 064020 (2007).
\bibitem{zcwu} S. W. Hawking, I. G. Moss, and J. M. Stewart, Phys. Rev. D {\bf 26}, 2681
(1982); Z.C. Wu, Phys. Rev. D {\bf 28}, 1898 (1983); I. G. Moss,
Phys. Rev. D {\bf 50}, 676 (1994); D. Langlois, K.-i. Maeda, and
D. Wands, Phys. Rev. Lett. {\bf 88}, 181301 (2002); J. J.
Blanco-Pillado, M. Bucher, S. Ghassemi, and F. Glanois, Phys. Rev.
D {\bf 69}, 103515 (2004); J. Garriga, A. H. Guth, and A.
Vilenkin, Phys. Rev. D {\bf 76}, 123512 (2007); S. Chang, M.
Kleban, and T. S. Levi, arXiv:0712.2261; A. Aguirre and M. C.
Johnson, arXiv:0712.3038.
\bibitem{israel01} W. Israel, Nuovo Cimento {\bf 44B}, 1, (1966); {\it ibid.} {\bf
48B}, 463(E) (1967); V. A. Berezin, V. A. Kuzmin, and I. I.
Tkachev, Phys. Rev. D {\bf 36}, 2919 (1987); R. Mansouri and M.
Khorrami, J. Math. Phys. {\bf 37}, 5672, (1996); S. Ansoldi, A.
Aurilia, R. Balbinot, and E. Spallucci, Class. Quantum Grav. {\bf
14}, 2727 (1997).
\bibitem{oppen01} R. B. Mann and S. F. Ross, Phys. Rev. D {\bf 47}, 3319
(1993); R. B. Mann and J. J. Oh, Phys. Rev. D {\bf 74}, 124016
(2006); S. Hyun, J. Jeong, W. Kim, and J. J. Oh, J. High Energy
Phys. 04 (2007) 088; S. Hyun, J. Jeong, W. Kim, and J. J. Oh,
Class. Quantum Grav. {\bf 24}, 3465 (2007).
\bibitem{oppen02} J. R. Oppenheimer and H. Snyder, Phys. Rev. {\bf 56}, 455
(1939).
\bibitem{hsato01} H. Sato, Prog. Theor. Phys. {\bf 76},
1250 (1986); A. Aurilia, M. Palmer, and E. Spallucci, Phys. Rev. D
{\bf 40}, 2511 (1989); C. Barrabes and W. Israel, Phys. Rev. D
 {\bf 43}, 1129 (1991); K. M. Larsen and R. L. Mallett, Phys. Rev.
D {\bf 44}, 333 (1991); N. Sakai and K.-i. Maeda, Phys. Rev. D
{\bf 50}, 5425 (1994); F. S. N. Lobo and P. Crawford, Class.
Quantum Grav. {\bf 22}, 4869 (2005); M. Gogberashvili, Phys. Lett.
B {\bf 636}, 147 (2006).
\bibitem{nam01} B.-H. Lee, W. Lee, S. Nam, and C. Park, Phys.
Rev. D {\bf 75}, 103506 (2007).
\bibitem{ran01} L. Randall and R. Sundrum, Phys. Rev. Lett. {\bf 83}, 3370
(1999); {\it ibid.} {\bf 83}, 4690 (1999).
\bibitem{kraus01} P. Kraus, J. High Energy Phys. 12 (1999) 011; C.
Park and S. J. Sin, Phys. Lett. B{\bf 485}, 239 (2000); H. Collins
and B. Holdom, Phys. Rev. D {\bf 62}, 105009 (2000); N. J. Kim, H.
W. Lee and Y. S. Myung, Phys. Lett. B{\bf 504}, 323 (2001); Y. S.
Myung, J. Korean Phys. Soc. {\bf 43}, 991 (2003).
\bibitem{cham01} H. A. Chamblin and H. S. Reall, Nucl. Phys. B{\bf
562}, 133 (1999).
\bibitem{barcelo01} C. Barcelo and M. Visser, Phys. Rev. D {\bf 63}, 024004
(2000).
\bibitem{ygh} J. W. York, Phys. Rev. Lett. {\bf 28}, 1082 (1972);
G. W. Gibbons and S. W. Hawking, Phys. Rev. D {\bf 15}, 2752
(1977).
\bibitem{misner} C. W. Misner, K. S. Thorne, and J. A. Wheeler, {\it
Gravitation} (Freeman, San Francisco, 1973).
\bibitem{birkhoff} G. D. Birkhoff, {\it
Relativity and Modern Physics} (Harvard University Press,
Cambridge, England, 1923).
\bibitem{ipser} J. Ipser and P. Sikivie, Phys. Rev. D {\bf 30}, 712 (1984).
\bibitem{boul1} D. G. Boulware, Phys. Rev. D {\bf 8}, 2363 (1973).
\bibitem{mann2} B. D. Miller, Astrophys. J. {\bf 208}, 275 (1976); R. Mann, Class. Quantum Grav. {\bf 14}, 2927
(1997); R.-G. Cai, J.-Y. Ji, and K.-S. Soh, Phys. Rev. D {\bf 57},
6547 (1998); G. T. Horowitz and R. C. Myers, Phys. Rev. D {\bf
59}, 026005 (1998); H.-a. Shinkai and T. Shiromizu, Phys. Rev. D
{\bf 62}, 024010 (2000); B. Freivogel, G. T. Horowitz, and S.
Shenker, J. High Energy Phys. 05 (2007) 090.
\bibitem{vil03} A. Vilenkin, Phys. Lett. B {\bf 117}, 25 (1982).
\bibitem{eternal} A. Vilenkin, Phys. Rev. D {\bf 27}, 2848 (1983); A. H. Guth, Phys. Rep. {\bf 333}, 555 (2000).

\end{thebibliography}
\end{document}